# Etching-free transfer of wafer-scale $MoS_2$ films

Donglin Ma[1], Jianping Shi[1,2], Qingqing Ji[1], Yu Zhang[1,2], Mengxi Liu[1], Qingliang Feng[1], Xiuju Song[1], Jin Zhang[1], Yanfeng Zhang[1,2*], Zhongfan Liu[1*]

[1]Center for Nanochemistry (CNC), Beijing National Laboratory for Molecular Sciences, State Key Laboratory for Structural Chemistry of Unstable and Stable Species, College of Chemistry and Molecular Engineering, Peking University, Beijing 100871, People's Republic of China
[2]Department of Materials Science and Engineering, College of Engineering, Peking University, Beijing 100871, People's Republic of China
E-mail: (yanfengzhang@pku.edu.cn, zfliu@pku.edu.cn)

**How to transfer $MoS_2$ films from growth substrates onto target substrates is a critical issue for its practical applications. However, it remains a great challenge to avoid the sample degradation and substrate destruction, since current transfer method inevitably employs a wet chemical etching process. Herein, we develop an etching-free transfer method for transferring wafer-scale $MoS_2$ films onto arbitrary substrates by using ultrasonication. Briefly, the collapse of ultrasonication-generated microbubbles at the interface between polymer-coated $MoS_2$ film and substrates induce sufficient force to delaminate the $MoS_2$ films. Using this method the $MoS_2$ films can be transferred from all the substrates (silica, mica, strontium titanate, sapphire) and remains the original sample morphology and quality. This method guarantees a simple transfer process, allows the reuse of growth substrates, without the presence of any hazardous etchants. The etching-free transfer method may promote the broad applications of $MoS_2$ in electronics, optoelectronics and catalysis.**

In recent years, atomically thin two-dimensional layered transition metal dichalcogenides (TMDCs) have attracted tremendous attention due to their intriguing electronic and optical properties[1-5]. One of the widely investigated compounds, monolayer molybdenum disulfide ($MoS_2$), possesses unique properties of tunable bandgap[6,7], strong light matter interaction[8], and strong spin-orbit coupling[9]. These properties make $MoS_2$ complementary to graphene[10,11], and therefore a perfect candidate for engineering a broad range of applications in field-effect transistors[12-15], photodetectors[16-18], and future spintronics and valleytronics[19-22].

Compared to the methods of mechanical exfoliation[12,23], liquid exfoliation[24], and solvothermal synthesis[25], chemical vapour deposition (CVD)[26-38] has been proved to be the most effective route to synthesize millimetre scale uniform monolayer $MoS_2$ on various substrates such as $SiO_2$ on Si ($SiO_2$/Si)[27,35-37], mica[33], strontium titanate (STO)[32], and sapphire[30,34,38]. A major challenge, however, is transferring CVD-grown $MoS_2$ onto target substrates for characterization and further device fabrication while maintaining high fidelity, keeping both the morphological and the physicochemical properties intact. Currently, the most widely adopted in $MoS_2$ transfer is a wet-etching transfer method in which polymethyl methacrylate (PMMA) is coated onto the $MoS_2$ as a film support and the composite stack is separated from the substrate using chemical etching. In this transfer process, which is in part an imitation of the transfer of graphene[39-42], chemical etching of the substrate[26,35,36,43] is inevitably involved, which leads to degraded film quality, as well as damage to or even consumption of the substrate. Particularly, in the etching of the insulating substrates for $MoS_2$ growth, treatments using hydrogen fluoride (HF) or strong alkali, which are harsher than the graphene etching treatments that use iron chloride ($FeCl_3$)[40] or ammonium persulfate (($NH_4$)$_2S_2O_8$)[42], present a larger corrosive hazard. In light of these factors, there would be great value in developing a "green" transfer method with high fidelity, high efficiency, recyclable use of substrates, and increased environmental friendliness[43,44].

In this work, a facile and etching-free transfer method is described for transferring $MoS_2$ from currently used insulating growth substrates to target substrates. This method is aided by an ultrasonic process, which generates millions of micron-sized cavitation bubbles. The collapse of these bubbles produces sufficient force in the interface between the PMMA-capped $MoS_2$ and the substrate to drive overlayer delamination. This transfer method, referred to as "ultrasonic bubbling transfer", is free of any chemical etchant and thus analogous to a physical exfoliation process; because it introduces no hazardous pollutants to the environment, it can be seen as a "green" alternative to traditional chemical etching. The initial crystal quality of the film can be maintained to a large extent, which makes it possible to investigate various properties of the transferred $MoS_2$. Additionally, perfect preservation of the flatness of the growth substrate means that the substrate can be reused for more growth cycles, thus dramatically reducing both waste and the production cost.

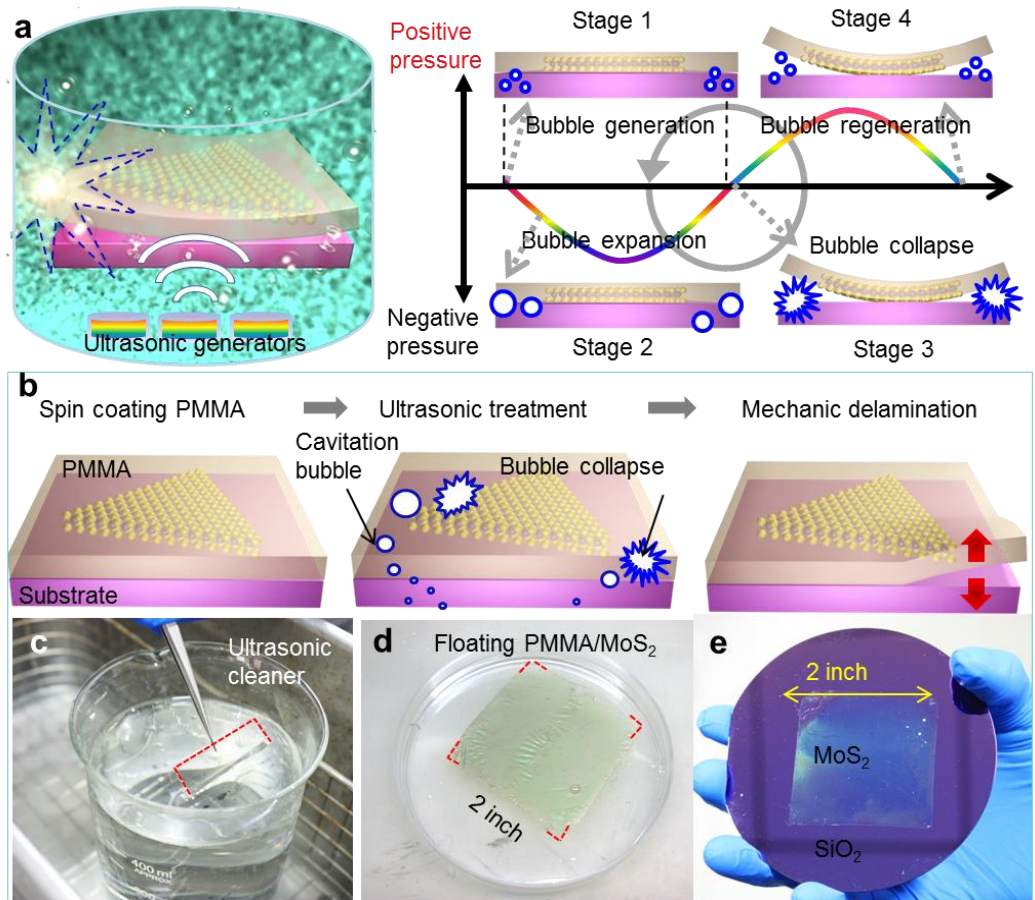

**Figure 1. Illustration of ultrasonic bubbling transfer of MoS$_2$.** (a) Schematics and principles of PMMA/MoS$_2$ stack delamination from the growth substrate. (b) Strategy for the ultrasonic bubbling transfer of MoS$_2$ on insulating substrates. (c) PMMA/MoS$_2$/Mica stack immersed into a beaker of water kept in an ultrasonic cleaner. (d) 2-inch by 2-inch PMMA/MoS$_2$ stack floating on water. (e) Large-area transfer of MoS$_2$ onto arbitrary substrates like SiO$_2$/Si.

**General ultrasonic bubbling transfer method**

Figure 1a shows a schematic illustration of the ultrasonic process used to detach a PMMA/MoS$_2$ stack from the growth substrate. A typical bubbling cycle is presented in the right panel of Figure 1a, with the middle sine curve showing the pressure change as a function of ultrasonication time. Under an ultrasonic cycle, millions of cavitation bubbles are generated at the initial of the negative pressure period of the ultrasonic wave (stage 1), and these bubbles rapidly expand into larger bubbles until the pressure suddenly switches from negative to positive (stage 2). In a short time, these bubbles are compressed and collapse at the rising of the positive pressure period (stage 3), releasing an enormous amount of energy, which produces considerable force in the interface between PMMA/MoS$_2$ stack and the insulating substrate. Within many bubbling cycles (stage 4), the bubbling-induced force steadily delaminates the PMMA/MoS$_2$ stack from the growth substrate.

Figure 1b illustrates the key steps in the transfer process. As-grown MoS$_2$ on the growth substrate (SiO$_2$/Si, mica, STO, or sapphire) is first spin-coated with PMMA, which is then cured. The PMMA/MoS$_2$/substrate stack is then immersed into a beaker of water kept in an ultrasonic cleaner (Figure 1c). Within one minute, the edge of the PMMA/MoS$_2$ film can be seen detaching from the substrate, eventually causing the PMMA/MoS$_2$ stack to float to the surface of the water (Figure 1d). The delaminated PMMA/MoS$_2$ films can then be transferred onto any target substrate (such as SiO$_2$, Figure 1e). Once the film is deposited on the target substrate, the PMMA can then be removed with acetone; this is the same process as the one used in graphene transfer. Notably, the entire delamination process uses only water, and involves no chemical etchants or hazardous pollutants.

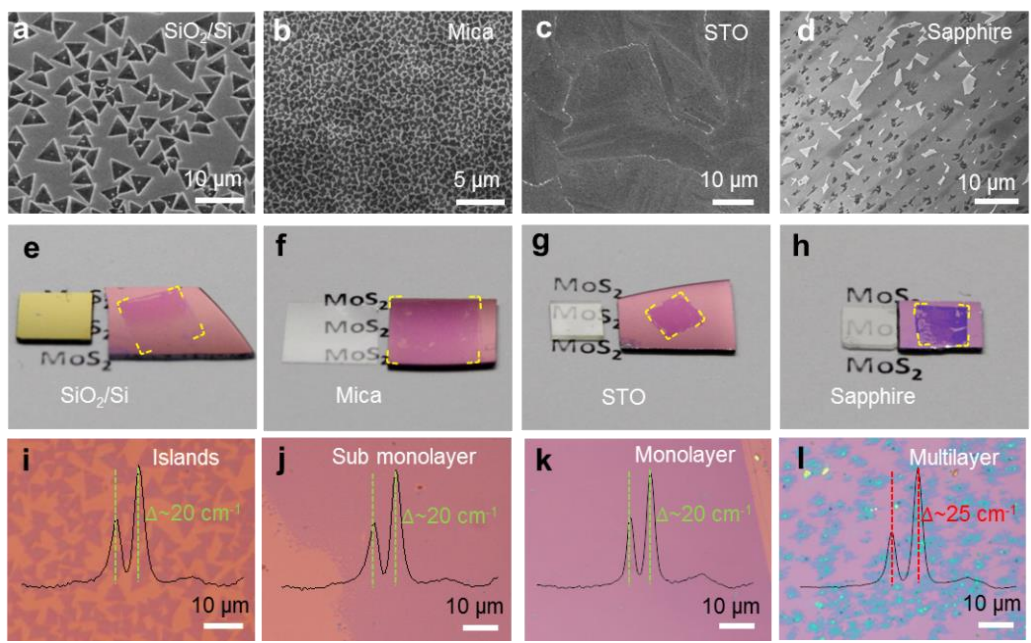

**Figure 2. Universal transfer of MoS$_2$ grown on SiO$_2$/Si (a, e, i), mica (b, f, j), STO (c, g, k) and sapphire (d, h, l).** (a–d) SEM images of as-grown MoS$_2$ on corresponding insulating substrates. (e–h) Digital photographs of the insulating substrates after sample transfer (left) and MoS$_2$ transferred onto SiO$_2$/Si target substrates (right). (i–l) Optical microscope images of MoS$_2$ transferred onto SiO$_2$/Si target substrates corresponding to (e–h), respectively. The overlaid spectra are corresponding Raman characterizations of the layer numbers ((i–k) for monolayer, and (l) for multilayer in the deep contrast regions).

**Transferring MoS$_2$ films from various growth substrates**

To demonstrate the universality of our transfer method, MoS$_2$ films were synthesized on SiO$_2$/Si, mica, STO, and sapphire growth substrates, and transferred to a target SiO$_2$/Si substrate. Different growth methods were also adopted. The sample grown on SiO$_2$/Si was prepared by physical vapour deposition (PVD)[45] while the rest were grown by CVD. The growth substrates were selected because they displayed variable film coverage, from submonolayer to monolayer and even multilayer (> 4 layers), and variable MoS$_2$ flake shapes, from individual triangular islands to merged films. These variations are shown in SEM images in Figure 2a–d. As shown in the photographs in Figure 2e–h, purple contrasts with the same dimensions as the growth substrates are clearly visible after sample transfer onto SiO$_2$/Si target substrates. The non-uniform coverage of the MoS$_2$ grown on SiO$_2$/Si transferred to the target substrate (Figure 1e) arises from the uneven growth of MoS$_2$ on SiO$_2$ from the PVD growth process (Figure S1). Corresponding optical microscope (OM) images (Figure 2i–l) of the transferred samples, which are shown on a comparable length scale to the SEM images, display a perfect preservation of the coverage and morphology of the MoS$_2$ films throughout the ultrasonic bubbling transfer process. Corresponding Raman characterizations show frequency differences of 20 cm$^{-1}$ and 25 cm$^{-1}$ between two main peaks in the spectrum (corresponding to Raman modes of E$^1_{2g}$, A$_{1g}$), which correspond to the monolayer and multilayer, respectively[46].

These results indicate that this ultrasonic bubbling transfer method can be widely used to transfer MoS$_2$ films synthesized on various substrates with perfect preservation of the coverage, morphology, and thickness of the MoS$_2$ film. It is noteworthy that this method is also effective in transferring WS$_2$[47], another widely used transition metal dichalcogenide (Figure S2), which suggests that the ultrasonic bubbling transfer method could be applied to more other transition metal dichalcogenides.

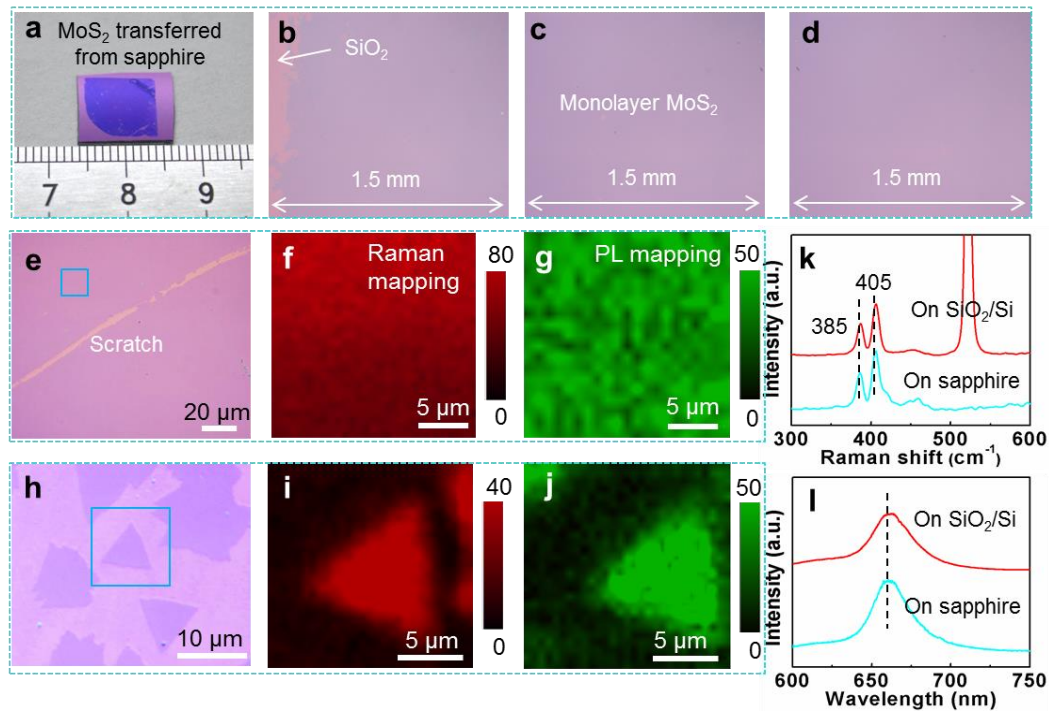

**Figure 3. Large-scale OM and Raman characterizations of MoS$_2$ transferred onto SiO$_2$/Si.** (a) Photograph showing the centimetre-scale uniform transfer of a monolayer MoS$_2$ film grown on sapphire onto a Si/SiO$_2$ substrate. (b–d) Sequential OM images of sample (a) captured from the left boundary to the inner part of the sample showing a uniform colour contrast. (e, h) OM images of a monolayer film consisting a scratch line and submonolayer triangular MoS$_2$ flakes, respectively. (f, i) Corresponding Raman mapping of the area marked by squares in (e, h), respectively. (g, j) PL mapping of the same areas of (f, i) showing uniform colour contrasts over the MoS$_2$ covered regions, indicative of almost no crystal quality degradation through the transfer process. (k, l) Raman and PL spectra collected from randomly chosen locations on the monolayer MoS$_2$ on the sapphire growth substrate and Si/SiO$_2$ target substrate.

In addition to preserving the film morphology, preserving the crystal quality is an essential parameter for evaluating a transfer method. Figure 3a shows the characterization of a monolayer MoS$_2$ film grown on sapphire and transferred onto a Si/SiO$_2$ target substrate. The colour photographic images show a uniform colour contrast within a sample size of around 1×1 cm. A series of OM images (Figure 3b–d) were captured in sequence from the left margin to the inner part of the transferred film, covering a total length of 4.5 mm. Uniform and continuous contrasts can be clearly seen, dispelling the concern that the bubbling process may have caused mechanical damage to the transferred MoS$_2$ films. In order to confirm the high fidelity of the crystal quality, Raman (from 370 to 420 cm$^{-1}$) and PL mappings (from 640 to 670 nm) were obtained for a monolayer film and a triangular island (OM images shown in Figure 3e, h) respectively. These results show uniform contrasts in the mapped regions (Figure 3f, i, Figure 3g, j), which are indicative of uniform crystal quality. Meanwhile, single-point Raman spectra from the transferred samples (Figure 3k) exhibit two typical peaks, with the in-plane vibration of Mo and S atoms (E$^1_{2g}$) at ~385 cm$^{-1}$ and the out-of-plane vibration of S atoms (A$_{1g}$) at ~405 cm$^{-1}$ with a frequency difference Δ ~20 cm$^{-1}$, which are in line with the Raman spectra of the pristine monolayer MoS$_2$ on the sapphire growth substrate. Moreover, single-point PL spectra (Figure 3l) show the intense characteristic peak of A excitonic emission at 662 nm for both the pristine and transferred MoS$_2$ samples, again suggesting perfect preservation of the high crystal quality. A quality survey of transferred MoS$_2$ samples grown on the other three substrates is provided in the supplementary of Figure S3. These spectroscopy measurements indicate that all the transferred MoS$_2$ samples retain their pre-transfer properties. Of special note here is that by using this newly developed ultrasonic bubbling transfer method, the crystal quality of the transferred samples should be at least comparable, but more likely superior to the crystal quality obtained using the wet chemical etching method.

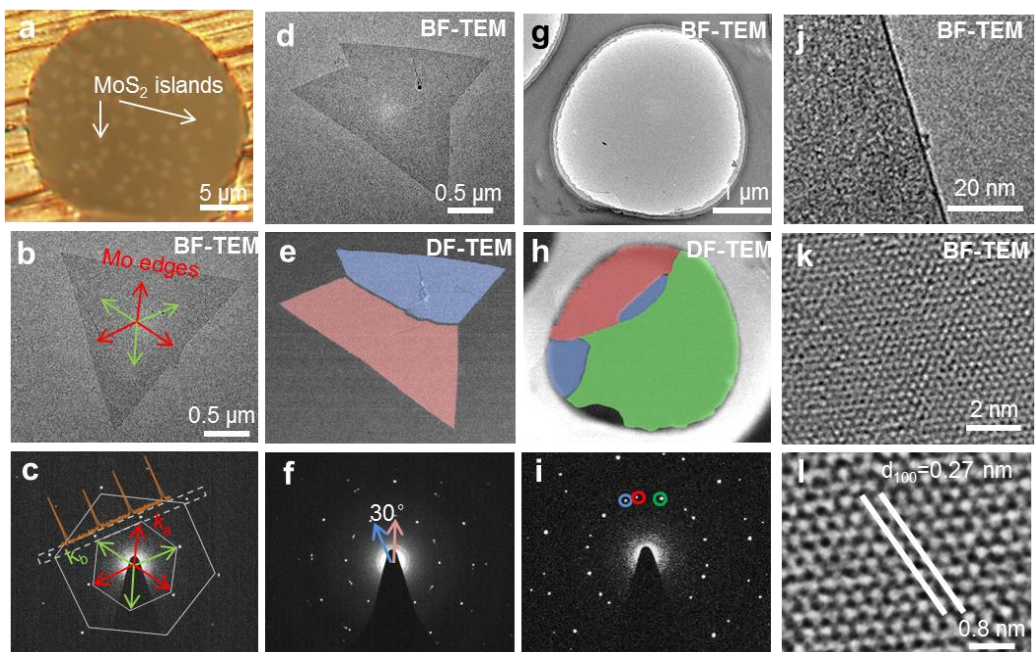

**Figure 4. Transferability of MoS$_2$ onto TEM grids *via* the ultrasonic bubbling transfer method.** (a) Optical image of multiple MoS$_2$ islands transferred onto carbon-film-coated Cu grids. (b) Bright-field TEM (BF-TEM) image of a MoS$_2$ triangle showing a Mo-zigzag edge orientation, as confirmed by the corresponding selected area electron diffraction (SAED) pattern shown in (c). The asymmetry of Mo and S sublattices divides the hexagonal diffraction spots into two families: $k_a$ and $k_b$. Overlaid is a line profile that was scanned along the white dashed line. The higher intensity $k_a$ spots refer to the Mo sublattice, as indicated by red arrows in (b). (d) BF-TEM image of a polycrystalline MoS$_2$ flake created by two merged triangular flakes. (e) False-colour dark-field TEM (DF-TEM) image corresponding to two sets of diffraction patterns in (f). (f) Diffraction pattern of the MoS$_2$ flake in (d); two sets of diffraction patterns are indicated by blue and pink arrows, respectively. (g) BF-TEM image of a continuous monolayer MoS$_2$ film transferred onto a lacey carbon-film-coated Cu TEM grid. (h) False-colour DF-TEM image corresponding to the three sets of diffraction patterns in (i). (j) Cross-section TEM image on the film edge. (k, l) Atomically resolved TEM images of monolayer MoS$_2$.

**Ultrasonic bubbling transfer for TEM sample preparation**

A significant advantage of this ultrasonic bubbling transfer method is that it can be utilized to prepare appropriate samples (similar to the one shown in Figure 3h) for TEM characterization. TEM is an essential characterization method for revealing the microscopic features of materials like MoS$_2$, where a reliable transfer process is inevitably involved to prepare appropriate samples. This allows for atomic-scale characterizations of such features of the film as the crystal quality, thickness, edge type, and grain boundary. The optical image in Figure 4a shows several triangle-shaped flakes of MoS$_2$ transferred onto the carbon-film-coated Cu grid. The morphology of the triangular MoS$_2$ flakes is well preserved through the transfer process. A typical triangular MoS$_2$ island with sharp edges is shown in Figure 4b, with corresponding selected area electron diffraction (SAED) (Figure 4c) showing diffraction spots with six-fold symmetry; this is highly suggestive of a single-crystal sample. In a detailed study, the diffraction spots in six-fold symmetry could be further divided into two sets of spots with three-fold symmetry, corresponding to the Mo and S sublattices. According to the line profile along the white dashed line (Figure 4c overlay), $k_a$ with higher intensity refers to the Mo sublattice with respect to $k_b$ with lower intensity in parallel with the S sublattice[35]. The relative intensity of $k_a$ and $k_b$ can be utilized to identify the edge type of a MoS$_2$ triangular flake (TEM image of single crystal MoS$_2$ islands with S-terminated edges in Figure S4). Besides single crystalline triangles, polycrystalline MoS$_2$ flakes formed by the aggregation of more than one triangle can also be perfectly transferred onto carbon-film-coated Cu grids by the ultrasonic bubbling transfer method. For example, Figure 4d shows the BF-TEM image of a typical polygonal-shaped MoS$_2$ island. The corresponding false-colour dark-field TEM (DF-TEM) image and SAED pattern in Figure 4e, f can be utilized to identify the domain boundary, and to observe a relative rotation of ~30° between the two composite domains.

In addition, as shown in Figure 4g, a uniform monolayer MoS$_2$ film was transferred intact onto lacey carbon-film-coated Cu grid, as evidenced by a freestanding MoS$_2$ film residing on a hole with a diameter of ~4 μm. In accordance with the diffraction spots in Figure 4i, the false-colour DF-TEM image in Figure 4h marks the different orientations of the MoS$_2$ domains with different colours. Moreover, a cross-section TEM image taken at the edge of the MoS$_2$ film shows only one dark line contrast. This provides definite proof that the transferred film is a monolayer. Of particular significance is the fact that high-resolution TEM images in Figure 4k and l exhibit perfect atomic lattices in hexagonal symmetry extending over tens of nanometres. The d(100) distance, as marked by two parallel white lines in Figure 4l, is measured to be around 0.27 nm, which is in agreement with the lattice constant of MoS$_2$. Hereby, the developed physical transfer method is qualified for improved TEM characterization of MoS$_2$ and similar materials.

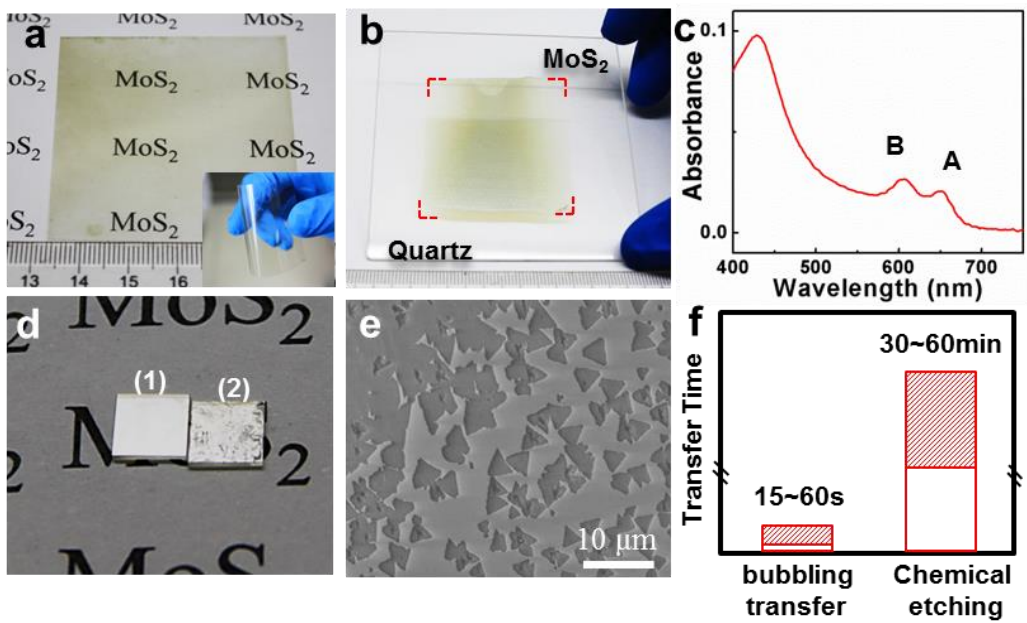

**Figure 5. Fast ultrasonic bubbling transfer of MoS$_2$ to recycle growth substrates.** (a) Wafer-scale (2-inch by 2-inch) MoS$_2$ film grown on mica *via* atmospheric pressure CVD. The inset shows a bent MoS$_2$ film on mica that displays perfect flexibility. (b) MoS$_2$ films transferred onto a quartz substrate *via* the ultrasonic bubbling method. (c) UV-Vis spectra of monolayer MoS$_2$ on quartz. (d) Photograph of sapphire

substrates after ultrasonic bubbling transfer (1) and after chemical etching transfer (2). (e) SEM image presenting the successful growth of MoS$_2$ on recycled sapphire (substrate 1) after several growth and transfer process. (f) Efficiency comparison of ultrasonic bubbling and chemical etching transfer processes.

The as-grown MoS$_2$ on mica serves as a good candidate for flexible electronics because of its perfect transparency and bendability (Figure 5a). As shown in Figure 5b, this ultrasonic bubbling transfer method can transfer entire, intact MoS$_2$ films onto 3-inch quartz plates. The UV-Vis spectrum in Figure 5c displays two absorption peaks at 653 nm (1.90 eV) and 606 nm (2.05 eV), with the peak positions and the corresponding energy difference (0.15 eV) in agreement with the theoretical value (0.148 eV) for monolayer MoS$_2$[48]. This demonstrates a perfect preservation of the optical (or the crystalline) quality of MoS$_2$ films after the ultrasonic bubbling transfer method, and the transferred film is of macroscopically uniform thickness. Field effect transistor (FET) devices were also fabricated to demonstrate the suitability of the bubbling transferred samples for electric devices (Figure S5). The extracted carrier mobility of MoS$_2$ FET devices were around 0.04 cm$^2$ V$^{-1}$ s$^{-1}$. It is likely that carefully selected electrodes and the use of a high-κ top gate material will greatly enhance the carrier mobility of the MoS$_2$ FET device.

**Enabling reuse of growth substrates**
Another important aspect of evaluating a transfer method is determining its ability to preserve the growth substrate for reuse. As the ultrasonic bubbling transfer method is similar to physical exfoliation, it was likely that this recycling would be possible. Figure 5d presents the surviving sapphire substrates from bubbling transfer (1) and chemical etching transfer (2). It is evident in the photographs that substrate 1 has retained its smooth surface, which indicates the pristine sapphire single crystal has not been damaged. In contrast, the substrate 2 appears to have a very rough surface. Substrate 1 was further characterized by atomic force microscopy (AFM), which showed the surface to be quite clean and free of any defects or residue (Figure S6).

Interestingly, triangular MoS$_2$ flakes can be successfully grown on substrate 1 even after several transfers using ultrasonic bubbling (Figure 5e). The resultant flakes are similar to those produced on a new sapphire substrate (Figure S7a). By contrast, substrate 2 was dirty and rough after chemical etching, and was decorated by irregularly shaped residues of MoS$_2$ or MoO$_3$ particles (Figure S7b, c), rendering it unsuitable for reuse. In addition to the fact that it preserves the growth substrate surface for reuse, a significant advantage of the ultrasonic bubbling transfer method is that the delamination process is very fast, taking on average less than 1 min to complete, which significantly improve the efficiency of transfer. The chemical etching process, on the other hand, takes more than 30 min to completely delaminate a PMMA/MoS$_2$ stack from the growth substrate, as determined from experimental statistics and reports in the published literature[49] (Figure 5f).

**Conclusion**
The ultrasonic bubbling transfer method we developed provides a brand new "green" method for MoS$_2$ transfer. This method was successfully applied to the four main growth substrates and probably extended to other systems. More importantly, the etching-free transfer process is facile and fast, and totally "green" for its free of any contamination to MoS$_2$ samples and its environmental friendliness. Furthermore, this method preserves the surface of the growth substrate and enables its reuse for the growth of new MoS$_2$ films. This work could not only facilitate the characterization of the intrinsic properties of TMDCs, but also accelerate the potential application of MoS$_2$ in a broad range of devices and technologies.

**Methods**
**MoS$_2$ growth on SiO$_2$/Si, mica, STO, sapphire.** MoS$_2$ grown on SiO$_2$/Si was prepared according to the procedure described in ref. 45. For MoS$_2$ grown on mica, STO and sapphire, the growth was performed inside a tube furnace (Lindberg/Blue M) equipped with 3-inch-diameter quartz tube. Sulfur powder (Alfa Aesar, purity 99.9%) was placed outside the hot zone and heated by heating belts at 135 ℃. $MoO_3$ powder (Alfa Aesar, purity 99.9%) and growth substrates (Hefei Kejing Material Technology Co., Ltd) were sequentially placed inside the hot zone of the tube furnace. In a typical growth, argon gas at a rate of 50 sccm was used as the carrier gas and growth time was 1 h. More detailed growth parameter were described in ref. 32,33,38.

**Bubbling transfer methodology.** PMMA was spin-coated (950K, ALLRESIST, AR-P 679.04) onto the $MoS_2$ film on the growth substrate ($SiO_2$/Si, STO, mica, or sapphire; typically 1×1 cm) at a relatively low speed of 1000–1500 rpm to yield a PMMA film with a thickness of 0.3 μm. The film was annealed at 180 ℃ for 15 min to completely remove residual solvent. The PMMA/$MoS_2$/substrate stack was immersed into a beaker of water that was kept in an ultrasonic cleaner (SB-3200 DTDN, Ningbo Scientz Biotech. Co., Ltd., 180W). After an optimized ultrasonication time (less than 1 min), the PMMA/$MoS_2$ stack delaminated from the substrate and was transferred onto the target substrate e.g. $SiO_2$/Si. Finally, the PMMA was removed with acetone.

**Characterization of transferred $MoS_2$.** The $MoS_2$ samples were systematically characterized using optical microscopy (Olympus DX51), SEM (Hitachi S-4800; acceleration voltage of 1-2 kV), Raman spectroscopy (Horiba, LabRAM HR-800), TEM (JEOL JEM-2100F LaB6; acceleration voltage of 200 kV), UV-vis-IR (Perkin-Elmer Lambda 950 spectrophotometer), AFM (Vecco Nanoscope III).

**Acknowledgements**
This work was financially supported by the National Natural Science Foundation of China (Grants 51222201, 51290272, 51472008, 51432002), the Ministry of Science and Technology of China (Grants 2012CB921404, 2013CB932603, 2012CB933404, 2011CB921903), and the Foundation for Innovative Research Groups of the National Natural Science Foundation of China (Grant 51121091).